\documentclass[sigconf,natbib=true]{acmart}

\usepackage{dcolumn,booktabs}
\usepackage{multirow}
\usepackage[flushleft]{threeparttable}
\usepackage{subfigure}
\usepackage{amsmath}
\usepackage{amsfonts}
\usepackage{bm}
\usepackage{amsthm}
\usepackage{enumitem}
\usepackage{algorithm}
\usepackage{algpseudocode}

\AtBeginDocument{%
  }

\copyrightyear{2024}
\acmYear{2024}
\acmDOI{XXXXXXX.XXXXXXX}

\setcopyright{acmlicensed}\acmConference[RecSys '24]{18th ACM Conference on Recommender Systems}{October 14--18, 2024}{Bari, Italy}
\acmBooktitle{18th ACM Conference on Recommender Systems (RecSys '24), October 14--18, 2024, Bari, Italy}

\acmISBN{978-1-4503-XXXX-X/18/06}

\begin{document}

\title{A Practice-Friendly LLM-Enhanced Paradigm with Preference Parsing for Sequential Recommendation}

\author{Dugang Liu}
\affiliation{%
  \institution{Guangdong Laboratory of Artificial Intelligence and Digital Economy (SZ)}
  \city{Shenzhen}
  \country{China}}  
\email{dugang.ldg@gmail.com}

\author{Shenxian Xian}
\affiliation{%
  \institution{CSSE, Shenzhen University}
  \city{Shenzhen}
  \country{China}
}
\email{xianshenxian2022@email.szu.edu.cn}

\author{Xiaolin Lin}
\affiliation{%
  \institution{CSSE, Shenzhen University}
  \city{Shenzhen}
  \country{China}
}
\email{linxiaolin2021@email.szu.edu.cn}

\author{Xiaolian Zhang}
\affiliation{%
  \institution{Huawei Technologies Co Ltd}
  \city{Shenzhen}
  \country{China}
}
\email{zhangxiaolian@huawei.com}

\author{Hong Zhu}
\affiliation{%
  \institution{Huawei Technologies Co Ltd}
  \city{Shenzhen}
  \country{China}
}
\email{zhuhong8@huawei.com}

\author{Yuan Fang}
\affiliation{%
  \institution{Huawei Technologies Co Ltd}
  \city{Shenzhen}
  \country{China}
}
\email{frank.fy@huawei.com}

\author{Zhen Chen}
\affiliation{%
  \institution{Huawei Technologies Co Ltd}
  \city{Shenzhen}
  \country{China}
}
\email{zzz.chen@huawei.com}

\author{Zhong Ming}
\authornote{Corresponding author}
\affiliation{%
  \institution{Guangdong Laboratory of Artificial Intelligence and Digital Economy (SZ)}
  \city{Shenzhen}
  \country{China}
}
\email{mingz@szu.edu.cn}

\renewcommand{\shortauthors}{Dugang Liu, et al.}

\begin{abstract}
The training paradigm integrating large language models (LLM) is gradually reshaping sequential recommender systems (SRS) and has shown promising results. However, most existing LLM-enhanced methods rely on rich textual information on the item side and instance-level supervised fine-tuning (SFT) to inject collaborative information into LLM, which is inefficient and limited in many applications. To alleviate these problems, this paper proposes a practice-friendly LLM-enhanced paradigm with preference parsing (P2Rec) for SRS. Specifically, in the information reconstruction stage, we design a new user-level SFT task for collaborative information injection with the assistance of a pre-trained SRS model, which is more efficient and compatible with limited text information. Our goal is to let LLM learn to reconstruct a corresponding prior preference distribution from each user's interaction sequence, where LLM needs to effectively parse the latent category of each item and the relationship between different items to accomplish this task. In the information augmentation stage, we feed each item into LLM to obtain a set of enhanced embeddings that combine collaborative information and LLM inference capabilities. These embeddings can then be used to help train various future SRS models. Finally, we verify the effectiveness and efficiency of our TSLRec on three SRS benchmark datasets.
\end{abstract}



\maketitle

\section{Introduction}\label{sec:intro}
Sequential recommender systems (SRS), which leverage historical interaction sequences to predict each user's next most likely item of interest~\cite{chang2021sequential,sun2019bert4rec,lin2024multi}, are core components in many industrial platforms.
A critical step in SRS is learning user preferences from interaction sequences better.
Previous studies have developed many technical routes for this purpose, the most representative of which include those based on recurrent neural networks (RNN)~\cite{donkers2017sequential,hidasi2015session,quadrana2017personalizing}, convolutional neural networks (CNN)~\cite{tang2018personalized,yuan2019simple,yan2019cosrec}, self-attention~\cite{kang2018self,he2021locker}, and multi-layer perceptron (MLP)~\cite{zhou2022filter,li2023automlp,gao2024smlp4rec}.
Recently, with the rapid development of large language models (LLM), the research of combining LLM with traditional SRS models to reshape the training paradigm has received increasing attention~\cite{zhao2024recommender}.

Existing LLM-enhanced SRS methods can be classified into three main categories: text information encoding-based, supervised fine-tuning (SFT) task-based, and structure distillation-based methods.
The first line focuses on using the rich textual information on the item side as a bridge between the LLM and SRS models. 
Then, it introduces various enhanced information embeddings encoded by LLM into the SRS model to improve recommendations~\cite{wang2024rethinking,yang2024sequential,geng2024breaking}.
This is also one of the most popular training paradigms.
The second line aims to design different supervised fine-tuning tasks, which are usually designed based on user interaction feedback, and inject the collaborative information required for recommendation into LLM and enable it to learn the preference mapping process required for recommendation, thus endowing it with the ability to act as a recommender for sequential recommender systems~\cite{li2023e4srec, zheng2024harnessing}.
The third line uses knowledge distillation to establish information alignment or transfer between LLM and SRS models at different layers, enhancing the SRS model's capabilities~\cite{li2023prompt, qiao2024llm4sbr}.

\begin{figure}[htbp]
    \centering
    \includegraphics[width=1.0\linewidth]{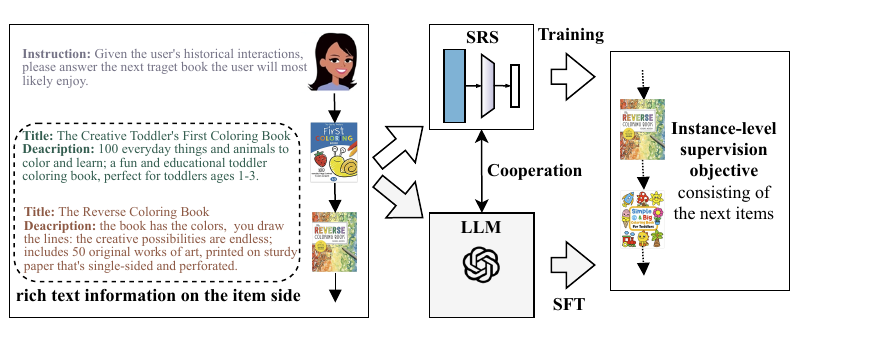}
    \caption{An illustration of a typical LLM-enhanced SRS model training architecture.}
    \label{fig:example}
\end{figure}

However, as shown in Fig.~\ref{fig:example}, the success of most existing LLM-enhanced methods is too dependent on rich item-related textual information and instance-level SFT tasks, which makes them inefficient and limited in many applications.
On the one hand, some scenarios may lack textual information related to items.
On the other hand, the number of interaction instances in industrial scenarios is usually very large, resulting in an excessively high and unacceptable LLM inference overhead for SFT.
To alleviate these problems, this paper proposes a practice-friendly LLM-enhanced paradigm with preference parsing (P2Rec) for SRS.
The core idea of our P2Rec is to introduce a new SFT task aimed at reconstructing users' prior preference distribution over all item latent categories, reasoning about the latent category of each item, and determining the relationship between different items. 
This reduces the reliance on textual information and the overhead of LLM from instance to user level.
Finally, extensive experiments are conducted on three public SRS benchmark datasets to answer the following questions: 1) How does our P2Rec perform compared to the baselines? 2) How does our P2Rec perform in terms of efficiency? 3) What is the gain of LLM for SRS in our P2Rec? 

\section{Related Work}\label{sec:related}
Sequential recommender systems (SRS) aim to capture user preferences from their historical interaction sequences and then use them to predict each user's next interaction item accurately.
As some of the most representative backbone models, FPMC~\cite{rendle2010factorizing}, GRU4Rec~\cite{hidasi2015session}, Caser~\cite{tang2018personalized}, SASRec~\cite{kang2018self}, and FMLP-Rec~\cite{zhou2022filter} adopt the form of Markov chain, recurrent neural network, convolutional neural network, self-attention, and multi-layer perception, respectively, to model the process of capturing user preferences.
Then, many follow-up works have been developed based on them to improve the effectiveness of this process further~\cite{chen2022intent,du2023ensemble}.
Recently, the great success of LLM in different tasks has attracted many efforts to apply LLM to SRS~\cite{wang2024rethinking,li2023e4srec,li2023prompt}.
However, most have limited applicability due to their instance-level SFT tasks and reliance on textual information.
Our P2Rec aims to provide a novel LLM-enhanced paradigm for SRS, enabling a better trade-off between performance and efficiency.

\section{Problem Formulation}\label{sec:problem}
In this section, we first define sequential recommendation with some necessary notations.
We denote the sets of users and items as $\mathcal{U}=\{u_{1},u_{2},\cdots,u_{M}\}$ and $\mathcal{V}=\left \{v_{1}, v_{2},\cdots,v_{N}\right \}$, respectively, where $M$ and $N$ represent the number of users and the number of items, respectively. 
Each user $u\in \mathcal{U}$ has an associated a sequence of historical interactions $\mathcal{S}^u=\left \{ v^u_1, v^u_2,\cdots,v^u_t,\cdots,v^u_{|\mathcal{S}^u|}\right \} $ and $\mathcal{S}^u$ is arranged by time, where $v^u_t\in\mathcal{V}$ denotes the item that the user interacts with at step $t$.
We then can formalize the sequential recommendation task.

\noindent\textbf{Input}: The set of historical interaction sequences for all users, i.e., $\mathcal{S}=\left \{ \mathcal{S}^{u_1}, \mathcal{S}^{u_2},\cdots,\mathcal{S}^{u_M} \right \} $.

\noindent\textbf{Output}: A SRS model to accurately predict an item $v$ that is most likely to be interacted by a user $u$ at the next moment.

\section{The Proposed Framework}\label{sec:method}
\begin{figure*}[htbp]
    \centering
    \includegraphics[width=0.85\textwidth]{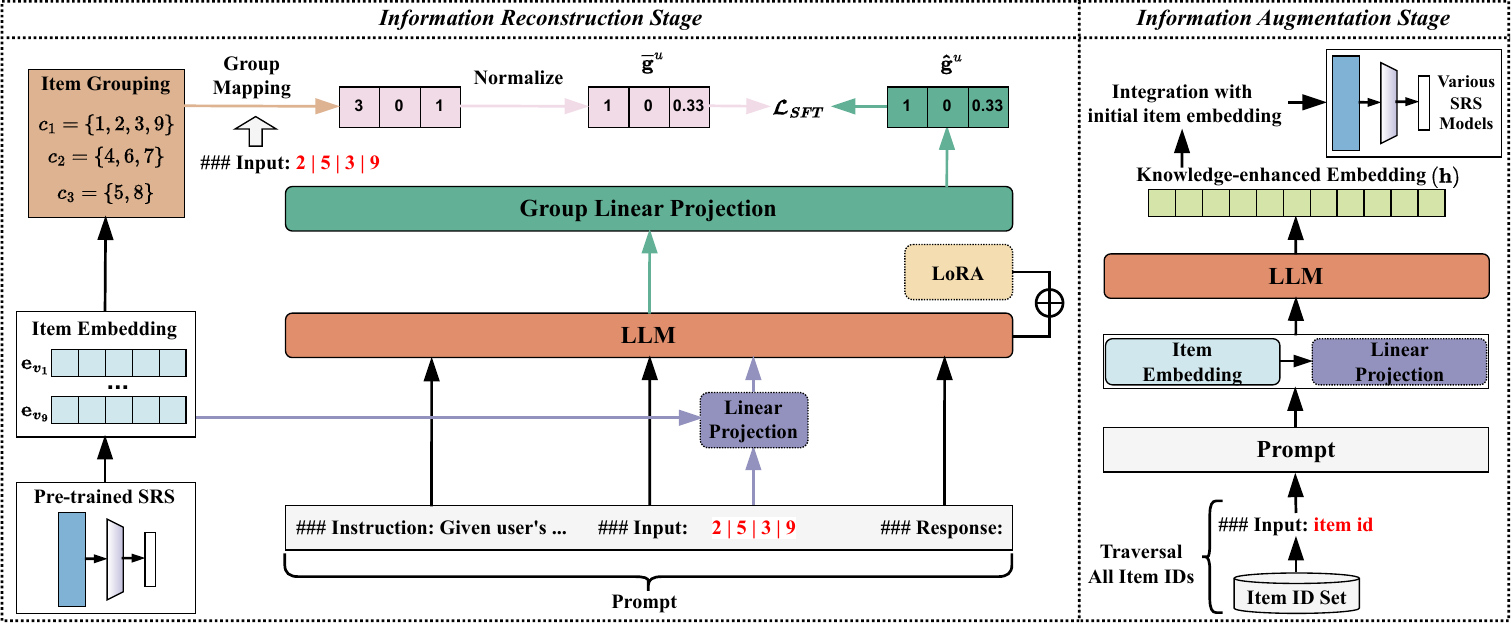}
    \caption{The architecture of our practice-friendly LLM-enhanced paradigm with preference parsing (P2Rec) framework.}
    \label{fig:framework}
\end{figure*}

In this section, we propose a novel practice-friendly LLM-enhanced paradigm with preference parsing (P2Rec), including an information reconstruction stage and an information augmentation stage, that is more attractive for deployment in industrial SRS.
Next, based on the training pipeline, we will describe these two core stages in detail.
To facilitate understanding, an illustration of our P2Rec framework is shown in Fig.~\ref{fig:framework}.

\subsection{Information Reconstruction Stage}
To avoid the overhead dilemma brought by instance-level SFT, we design a new user-level task for SFT at this stage, aiming to reconstruct the user’s prior preference distribution over all item latent categories.
This means that using LLM in our P2Rec will have lower complexity, i.e., reducing from $O(\sum_{i=1}^M|\mathcal{S}^{u_i}|)$ to $O(M)$.

\subsubsection{Characterize User Preferences}
\begin{figure}[htbp]
    \centering
    \includegraphics[width=0.95\linewidth]{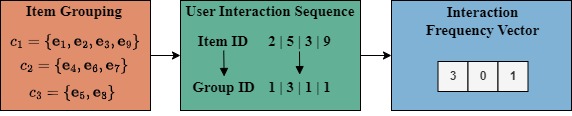}
    \caption{Illustration of the process of characterizing user preferences.}
    \label{fig:case1}
\end{figure}

To be compatible with more practical situations, we limit the side information of items to be unavailable and then utilize the collaborative knowledge of a pre-trained SRS model to obtain preliminary latent categories related to the items.
Specifically, given a backbone SRS model $M(\Theta)$ trained based on the ID paradigm, we can obtain a set of well-trained item embeddings from it, i.e., $\{\mathbf{e}_{v_1},\mathbf{e}_{v_2},\cdots,\mathbf{e}_{v_N}\}$.
We then perform a pre-grouping operation on them to obtain $K$ groups $\{c_1, c_2, \cdots, c_K\}$ and consider the group index corresponding to each item as its preliminary latent category.
For simplicity, we implement this operation using a $k$-means method~\cite{hartigan1979algorithm} in our experiments, which can be defined as follows:
\begin{equation}\label{eq:group}
    \{c_1, c_2, \cdots, c_K\}\leftarrow kmeans(\{\mathbf{e}_{v_1},\mathbf{e}_{v_2},\cdots,\mathbf{e}_{v_N}\}).
\end{equation}
After obtaining the mapping relationship between each item and its group, $c_i=C(\mathbf{e}_{v_j})$, where $C(\cdot)$ denotes the group mapping function and $c_i$ denotes the group corresponding to the $j$-th item, we can get the interaction frequency vector $\mathbf{g}^u=\left[g^u_1,g^u_2,\cdots,g^u_K\right]$ of each user in $K$ groups based on the items included in each user's interaction sequence, i.e., if user $u$ interacts with item $v$ and $v\in c_i$, $g^u_i$ will be increased by one; otherwise, it will remain at the initial value (i.e., 0).
In other words, the value of $g^u_i$ reflects the number of interactions that user $u$ has had with the set of items in the $i$-th group $c_i$.
A schematic of this process is shown in Fig.~\ref{fig:case1} for ease of understanding.
To avoid magnitude differences between users with different activity levels, we will also normalize these vectors (denoted as $\overline{\mathbf{g}}^u$) and use them to characterize user preferences.

\subsubsection{SFT for Reconstructing User Preferences}
Next, we modify the SFT task to take a user's interaction sequence $\mathcal{S}^u$ as input and reconstruct the user's corresponding preference, i.e., the normalized interaction frequency vector $\overline{\mathbf{g}}^u$.
Specifically, for ease of description, we denote the prompt template for this task as three parts, i.e., $d_{inst}$: ``Given the user's purchase history, predict the user's distribution among $K$ purchase preferences.'', and $d_{ids}$: $\mathcal{S}^u$, $d_{res}$: ``Response:''.
Similar to previous work~\cite{li2023e4srec}, we use the item embeddings provided by the pre-trained backbone SRS model to replace LLM’s token embeddings for item IDs in interaction sequences, making LLM aware of the item correlations behind different IDs.
Therefore, obtaining LLM knowledge embedding $\mathbf{h}^u$ based on this prompt can be expressed as follows.
\begin{equation}\label{eq:1}
\mathbf{h}^u=LLM\left(\mathbf{p}_{inst};linear\_proj([\mathbf{e}_{v^u_1}, \mathbf{e}_{v^u_2},\cdots,\mathbf{e}_{v^u_{|\mathcal{S}^u|}}]);\mathbf{p}_{res}\right),
\end{equation}
where $\mathbf{p}_{inst}$ and $\mathbf{p}_{res}$ denote the token sequence embeddings obtained by LLM for $d_{inst}$ and $d_{res}$, and $linear\_proj(\cdot)$ denotes the operation of projecting each item embedding into the same dimension as the token embedding of LLM.
Finally, we equip the knowledge embedding $\mathbf{h}^u$ with a group linear projection layer to generate the normalized interaction frequency vector, i.e.,
\begin{equation}\label{eq:2}
\hat{\mathbf{g}}^u=softmax\left(\mathbf{W}_{gp}\mathbf{h}^u+\mathbf{b}_{gp}\right),
\end{equation}
where $\mathbf{W}_{gp}$ and $\mathbf{b}_{gp}$ are the learnable weight matrix and vector, and $softmax(\cdot)$ is the softmax operation.
We use the mean squared error as a constraint to force the LLM to reconstruct the target better and adopt LoRA~\cite{hu2021lora} as the fine-tuning paradigm.
\begin{equation}\label{eq:3}
    \mathcal{L}_{SFT}=\sum_{u=1}^M\sum_{i=1}^K(\overline{\mathbf{g}}^u_i-\hat{\mathbf{g}}^u_i)^2.
\end{equation}
The idea behind this process is that we hope to guide LLM in learning to reconstruct the user's prior preference distribution from the user interaction sequence through a set of user-level examples.
To complete this task, LLM must parse each item's latent category and the relationship between items from the preference distribution. 
This enhances each item's potential category to a certain extent, which will be used in the next stage.

\subsection{Information Augmentation Stage}
As mentioned above, to accurately reconstruct the normalized frequency interaction vector $\overline{\mathbf{g}}^u$, LLM must try to infer the latent category to which each item belongs. 
With the power of LLM, we expect it to recognize the reasonable part of the pre-grouping results (i.e., give the same latent category) and correct the unreasonable part (i.e., provide different latent categories).
The knowledge-enhanced embeddings associated with this inference process will then likely better characterize each item's properties and the connections between different items.
Next, we describe using the LLM trained with SFT to obtain knowledge-enhanced embeddings and inferred latent categories for each item.
Specifically, this can be easily achieved by replacing the user interaction sequence $\mathcal{S}^u$ in the previous stage with each item ID $v_i$ as $d_{ids}$ in turn. 
This can also be viewed as there are $N$ virtual user interaction sequences, each of which includes only one item.
In this case, the embedding obtained by Eq.\eqref{eq:1} will be regarded as the knowledge-enhanced embedding of each item, and the vector obtained by Eq.\eqref{eq:2} reflects the latent category probability distribution of each item.
Finally, these knowledge-enhanced embeddings can be directly used to train various future SRS models. 
To demonstrate the effectiveness of our P2Rec, especially the knowledge-enhanced embeddings obtained, in our experiments, we adopt a simple gating mechanism to combine them with randomly initialized item embeddings of a new backbone SRS model to represent each item and use them for subsequent training.
We leave exploring more diverse fusion methods as future work, and an example of this stage can be found on the right side of Fig.~\ref{fig:framework}.
\section{Experiments}\label{sec:experiments}

\subsection{Experiment Settings}

\textbf{Dataset and Evaluation Metrics.}
To evaluate the effectiveness of our P2Rec, following the setup of previous works~\cite{li2023e4srec,xu2024fairly}, we select three public benchmark datasets in our experiments: Beauty, MovieLens-1M, and Yelp\footnote{https://github.com/RUCAIBox/RecSysDatasets}, and use the last interactive item, the penultimate item, and the remaining ones of each user for testing, validation, and training.
We summarize the statistics of the three processed datasets in Table~\ref{tab:dataset}.
Two widely used evaluation metrics will be adopted: hit ratio (HR@$k$) and normalized discounted cumulative gain (NDCG@$k$).
We report the average metrics for all users in the testing set, where $k$ is set to 5 and 10, respectively.

\begin{table}[htbp]
  \centering
  \caption{Statistics of the processed datasets.}
  \resizebox{1.0\linewidth}{!}{
    \begin{tabular}{c|ccccc}
    \toprule
    Dataset   & \#Users      & \#Items    & \#Interactions  & Density & Avg.Length \\
    \midrule
    Beauty   & 22,363     & 12,101    &198,502    & 0.07\%  & 8.9\\
    MovieLens-1M & 6,040    &3,416  &999,611    &4.84\% &165.5\\
    Yelp    & 30,431    &20,033 &316,354    &0.05\% &10.4\\
    \bottomrule
    \end{tabular}}
  \label{tab:dataset}
\end{table}

\begin{table*}[htbp]
\centering
\caption{Results on all datasets, where the best and second best results are marked in bold and underlined, respectively. Note that $^{*}$ indicates a significance level of $p\leq 0.05$ based on a two-sample t-test between our method and the best baseline.}
\resizebox{1.0\textwidth}{!}{
\begin{tabular}{l|l|cccc|cccc|cccc}
\toprule   
        & \multirow{2}{*}{Method} & \multicolumn{4}{c}{Beauty} & \multicolumn{4}{|c}{MovieLens-1M} & \multicolumn{4}{|c}{Yelp} \\
    \cline{3-14}
        & & \multicolumn{2}{c}{HR@\{5,10\}$\uparrow$} & \multicolumn{2}{c|}{NDCG@\{5,10\}$\uparrow$} & \multicolumn{2}{c}{HR@\{5,10\}$\uparrow$} & \multicolumn{2}{c|}{NDCG@\{5,10\}$\uparrow$} & \multicolumn{2}{c}{HR@\{5,10\}$\uparrow$} & \multicolumn{2}{c}{NDCG@\{5,10\}$\uparrow$} \\
\midrule
\multirow{4}{*}{LLM} 
           & POD  & 0.0185 & 0.0245 & 0.0125 & 0.0146 & 0.0422 & 0.0528 & 0.0291 & 0.0326 & 0.0476 & 0.0564 & 0.0330 & 0.0358\\
           & P5   & 0.0569 & 0.0791 & 0.0403 & 0.0474 & 0.2225 & 0.3131 & 0.1570 & 0.1861 & 0.0289 & 0.0453 & 0.0200 & 0.0252\\
           & LlamaRec & \underline{0.0591} & \underline{0.0862} & \underline{0.0405} & \underline{0.0492} & 0.1757 & 0.2836 & 0.1113 & 0.1461 & 0.0416 & 0.0605 & 0.0306 & 0.0367\\
           & E4SRec & 0.0527 & 0.0753 & 0.0376 & 0.0448 & 0.1871 & 0.2765 & 0.1234 & 0.1522 & 0.0309 & 0.0473 & 0.0207 & 0.0260\\
\midrule
\midrule
\multirow{3}{*}{GRU4Rec} 
           & +Base   & 0.0369 & 0.0590 & 0.0267 & 0.0329 & 0.2126 & 0.2912 & 0.1451 & 0.1703 & 0.0201 & 0.0365 & 0.0129 & 0.0182\\
           & +CI   & 0.0415 & 0.0603 & 0.0274 & 0.0335 & 0.2131 & 0.2957 & 0.1467 & 0.1734 & 0.0224 & 0.0368 & 0.0143 & 0.0189\\
           & +P2Rec   & 0.0442 & 0.0648 & 0.0299 & 0.0366 & 0.2147 & 0.3020 & 0.1498 & 0.1781 & 0.0242 & 0.0391 & 0.0152 & 0.0200\\
\midrule
\multirow{3}{*}{Caser} 
           & +Base   & 0.0259 & 0.0429 & 0.0162 & 0.0217 & 0.1902 & 0.2700 & 0.1282 & 0.1539 & 0.0181 & 0.0325 & 0.0113 & 0.0159  \\
           & +CI   & 0.0264 & 0.0433 & 0.0169 & 0.0223 & 0.1894 & 0.2722 & 0.1293 & 0.1558 & 0.0192 & 0.0350 & 0.0120 & 0.0166   \\
           & +P2Rec   & 0.0300 & 0.0494 & 0.0187 & 0.0249 & 0.1949 & 0.2762 & 0.1331 & 0.1593 & 0.0215 & 0.0366 & 0.0135 & 0.0183 \\
\midrule
\multirow{3}{*}{SASRec} 
           & +Base   & 0.0551 & 0.0846 & 0.0329 & 0.0424 & \underline{0.2364} & 0.3217 & \underline{0.1631} & 0.1907 & 0.0449 & 0.0637 & 0.0334 & 0.0394  \\
           & +CI   & 0.0569 & 0.0836 & 0.0354 & 0.0440  & 0.2315 & \underline{0.3257} & 0.1619 & \underline{0.1924} & 0.0467 & 0.0667 & 0.0344 & 0.0408\\
           & +P2Rec   & 0.0585 & 0.0854 & 0.0371 & 0.0458 & $\textbf{0.2404}^{*}$ & $\textbf{0.3315}^{*}$ & $\textbf{0.1670}^{*}$ & $\textbf{0.1965}^{*}$  & 0.0475 & 0.0685 & 0.0346 & 0.0414\\
\midrule
\multirow{3}{*}{FMLP-Rec} 
           & +Base   & 0.0577 & 0.0866 & 0.0361 & 0.0455  & 0.2230 & 0.3172 & 0.1545 & 0.1848 & 0.0491 & 0.0698 & 0.0356 & 0.0422 \\
           & +CI   & 0.0575 & \textbf{0.0868} & 0.0365 & 0.0459  & 0.2273 & 0.3123 & 0.1586 & 0.1861 & \underline{0.0511} & \underline{0.0743} & \underline{0.0365} & \underline{0.0439} \\
           & +P2Rec  & $\textbf{0.0604}^{*}$ & 0.0852 & $\textbf{0.0445}^{*}$ & $\textbf{0.0509}^{*}$  & 0.2341 & 0.3192 & 0.1626 & 0.1901  & $\textbf{0.0531}^{*}$ & $\textbf{0.0771}^{*}$ & \textbf{0.0373} & $\textbf{0.0451}^{*}$\\
\bottomrule
\end{tabular}}
\label{tab:main_result}
\end{table*}

\noindent\textbf{Baselines and Implementation Details.}
We adopt four SRS methods with different representative architectures as backbone models to verify the compatibility of our P2Rec: 
GRU4Rec~\cite{hidasi2015session}, Caser~\cite{tang2018personalized}, SASRec~\cite{kang2018self}, and FMLP-Rec~\cite{zhou2022filter}.
We also select four typical LLM-enhanced methods for comparison: POD~\cite{li2023prompt}, P5~\cite{hua2023index}, LlamaRec~\cite{yue2023llamarec} and E4SRec~\cite{li2023e4srec}.
We adopt the implementation in the RecBole~\cite{zhao2021recbole} framework for SRS baselines and their own open-source repositories for LLM-enhanced baselines.
Llama2-13B is taken as the backbone large language model.
For a fair comparison, we set the learning rate to $1e^{-4}$, the batch size to 1024, the embedding dimension to 256, and using the Adam as optimizer.
We carefully tune the unique parameters of each baseline according to the suggestions in the original papers.
We also use an early stopping strategy, with the patience set to 10 times, to obtain the best model based on the NDCG@10 on the validation set.
We will make the source codes publicly available once the paper has been accepted.

\subsection{RQ1: Performance Comparison}\label{sec:experiments:rq1}
In addition to all baselines, to better evaluate the advantage of our P2Rec in inferring and utilizing the latent category information of items, we additionally introduce a variant (denoted as `CI') that takes the pre-grouped index of each item as a feature and initializes a learnable embedding for it in the backbone model.
We report the comparison results in Table~\ref{tab:main_result}.
We can have the following observations: 
1) Our P2Rec outperforms the base backbone model and the variant using pre-grouping results in all cases. This shows the effectiveness of our proposed SFT task and the knowledge-enhanced embeddings obtained by combining the capabilities of LLM.
2) The performance of existing LLM-enhanced baselines is dataset-dependent and does not necessarily have an advantage over traditional methods. However, our P2Rec achieves a stable gain on all datasets.


\subsection{RQ2: Efficiency Analysis}\label{sec:experiments:rq2}
\begin{figure}[htbp]
\centering
\includegraphics[width=1.\linewidth]{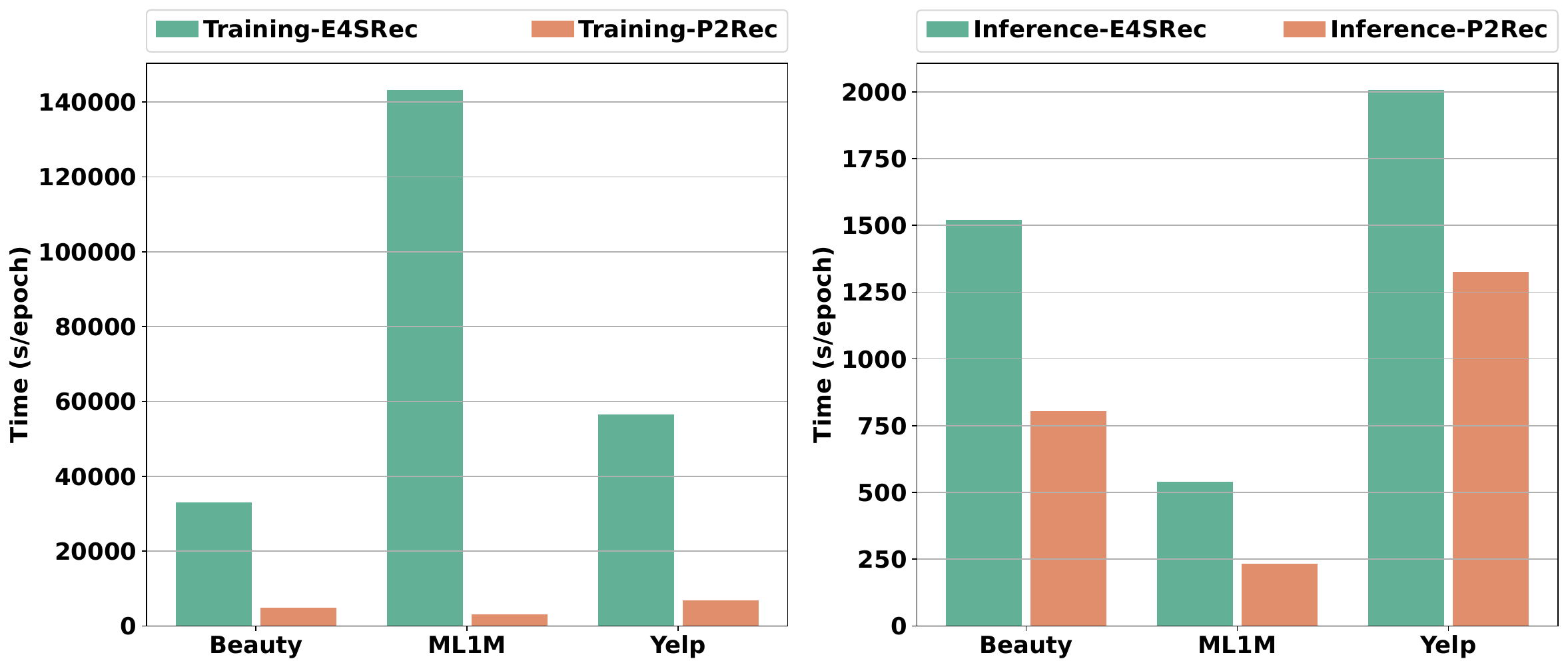}
\caption{The time required to execute LLM in one epoch in the training and inference phases of P2Rec and E4SRec, respectively.}
\label{fig:time}
\end{figure}

To verify the efficiency of our P2Rec, we take a recent LLM-enhance baseline, E4SRec, as an example and show the time it and our P2Rec take to perform LLM in training and inference phases, respectively.
The results are shown in Fig.~\ref{fig:time}, and comparison with other LLM-enhanced baselines has similar results.
Since our P2Rec modifies the complexity of SFT to the user level, we can find that its time to perform LLM in the training phase will be significantly reduced, especially on MovieLens-1M, which has the most training instances and the smallest number of users.
In addition, since our P2Rec only needs to obtain the knowledge-enhanced embedding of each item in the inference phase instead of using each user interaction sequence for inferencing like E4SRec, it also reduces the overhead.
This efficient property of P2Rec makes it more attractive for industrial deployment.

\subsection{RQ3: In-depth Analysis of P2Rec}\label{sec:experiments:rq3}
\begin{figure}[htbp]
\centering
\includegraphics[width=1.\linewidth]{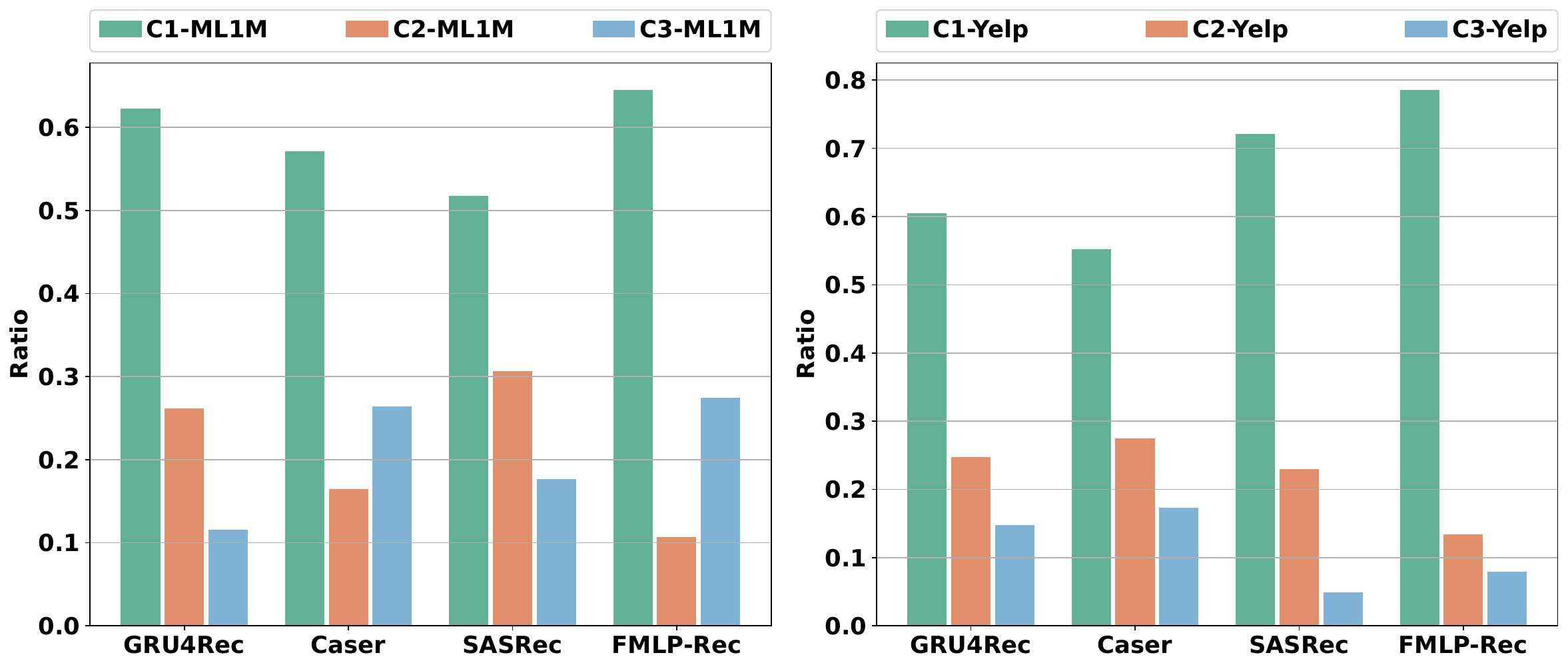}
\caption{The ratio of different comparison cases between the latent classes obtained by our P2Rec and the pre-grouped results.}
\label{fig:ratio}
\end{figure}

Finally, we provide some preliminary insights into how our P2Rec can leverage LLM's capabilities to enhance the training of SRS models.
We first compare the latent category vectors inferred for each item by the well-trained LLM via SFT with their corresponding pre-grouped indices.
The comparison results include three cases: 1) the category with the highest probability in the inferred vector is the same as the pre-grouping index (denoted as `C1'); 2) the categories with the top three probabilities include the pre-grouping index (denoted as `C2'); 3) others (denoted as `C3').
They respectively mean that the latent categories obtained using LLM in P2Rec agree with the pre-grouping result, think that an item belongs to multiple categories (including the pre-grouping result), or disagree with the pre-grouping result.
The results are shown in Fig.~\ref{fig:ratio}. 
We can see that our P2Rec revised the latent categories of some items based on the pre-grouping results, which may be an essential source of the gain of our P2Rec.
In other words, the LLM in our P2Rec can better identify each item's properties with the help of the designed SFT, which establishes the connections between different items.

\begin{figure}[htbp]
\centering
\includegraphics[width=1.\linewidth]{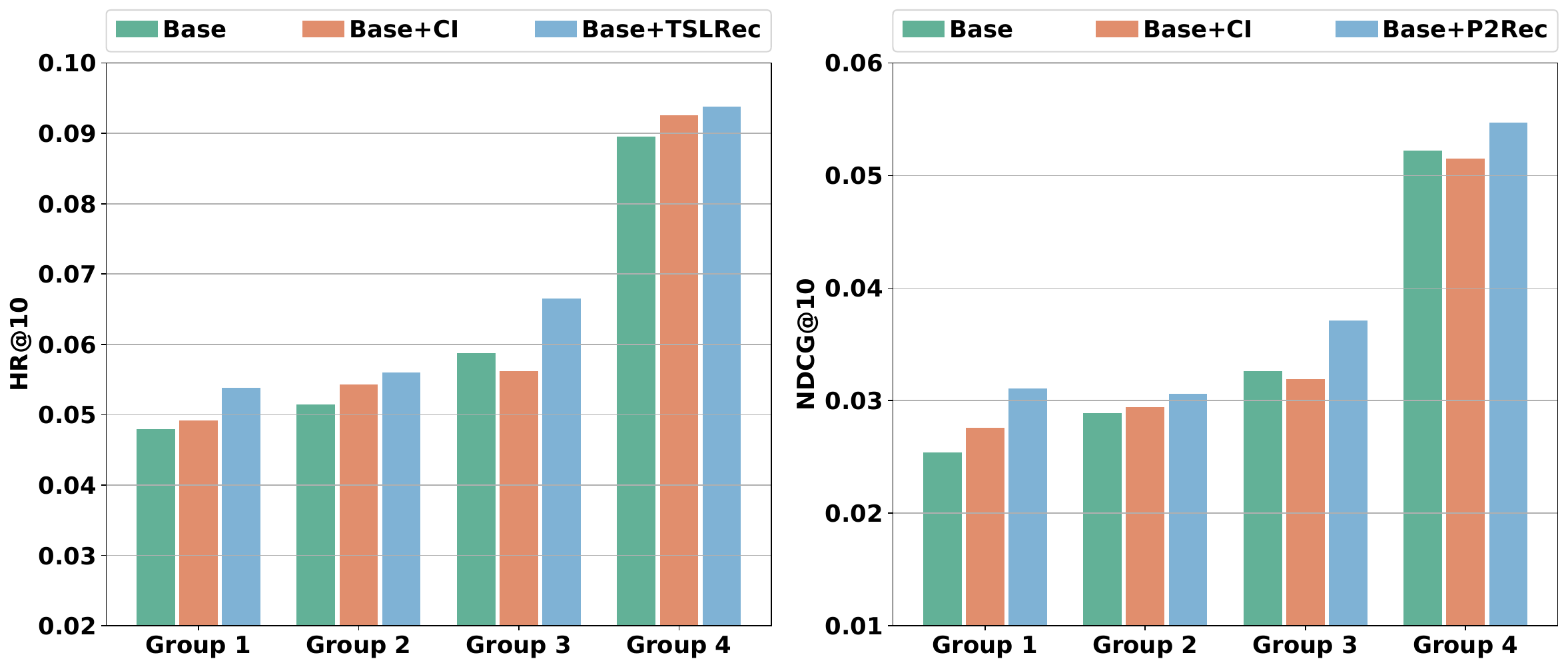}
\caption{The performance results of our P2Rec and baselines on different user groups.}
\label{fig:group}
\end{figure}

Second, we group each user by activity level and sort them in ascending order. 
As shown in Fig.~\ref{fig:group}, we report the results of our P2Rec and other baselines on all groups.
We can find that the variant that directly uses the pre-grouping results (i.e., `CI') may encounter performance bottlenecks in groups with high activity while our P2Rec consistently gains.
This may be because the variant `CI' uses the same feature embedding for all items in a group, which can easily confuse, while our P2Rec can provide more fine-grained enhanced embeddings for each item.

\section{Conclusions}\label{sec:conclusions}
In this paper, we consider the limitations of most existing sequential recommendation models augmented with large language models (LLM) regarding efficiency and rich textual dependencies and propose a novel practice-friendly LLM-enhanced paradigm with preference parsing (P2Rec).
Our P2Rec designs a novel supervised fine-tuning task for LLM in the first stage to reconstruct user preferences and reduce the training complexity to the user level.
In the second stage, we obtain an enhanced and informative embedding for each item by training a well-trained LLM, which can be directly used for training various future sequence models.
Finally, extensive experiments verify the effectiveness and efficiency of our P2Rec.

\bibliographystyle{ACM-Reference-Format}
\bibliography{sample-base}

\end{document}